\documentclass[reprint,amsmath,amssymb,aps,pra,prarmp,
prstab,
prstper,
floatfix]{revtex4-2}
\usepackage{color}
\usepackage{overpic}
\usepackage{graphicx}
\usepackage{dcolumn}
\usepackage{bm,ulem,soul}
\usepackage[colorlinks,citecolor=blue,linkcolor=red]{hyperref}
\date{\today}
\begin{document}
\title{Unbounded Work Extraction and Zero Work Fluctuations in a Super-Carnot Otto Information Engine}
\author{Yang Xiao$^{1}$}
\author{Jin Wang$^{2}$}\email{jin.wang.1@stonybrook.edu}
\affiliation{ $^1\,$College of Physics, Jilin University, Changchun 130022, China\\  $^2\,$ Department of Chemistry and Department of Physics and Astronomy, State University of New York at Stony Brook, Stony Brook, New York 11794, USA}

\begin{abstract} 
Generally, the work output of stochastic heat engines is governed by the stochastic trajectory distribution and the energy spectrum. Because the trajectory distribution depends on the thermal reservoir temperatures, the work output is not only constrained by temperature but is also susceptible to thermal fluctuations. To address these limitations, we introduce two continuous Maxwell's demons into a quantum Otto cycle, forming an Otto information engine (OIE). We demonstrate that the work output of the OIE depends solely on the energy level gap, enabling arbitrary  work extraction while eliminating  work fluctuations, thereby ensuring  cycle-to-cycle identical work output. Furthermore, we show that the engine's efficiency can surpass the standard Carnot efficiency even after accounting for the energy cost of demon's memory erasure. Finally, we show  that the OIE can deliver superior output power even when the demon's measurement time is  taken into account, and the corresponding Monte Carlo simulation has been executed. 
\end{abstract}
\maketitle
\date{\today}

\section{Introduction}
Heat engines \cite{MJ92,CR15,kos06,Quan07,GB17,NM22,SV16,Kosloff2014,VH18,Esposito10} are the  devices driven by a hot reservoir and a cold reservoir that convert heat into useful work. Unlike macroscopic classical engines that power modern industry \cite{YA01}, microscopic  heat engines have emerged as a pivotal platform for testing  thermodynamic laws \cite{MJ18,VH18,YX23,Seifert19,SS21}, exploring how to surpass classical limits by exploiting quantum resources, such as quantum coherence \cite{xiao23,TG19,MO03,Koslof15}, quantum entanglement \cite{YX23,lutz09}, reservoir squeezing \cite{lutz14,xiao23,JK17}, and quantum measurement \cite{Su21,CE17,CE18}, and active matter \cite{AD22,VH20,TE20}. Ultimately, these  engines offer fundamental design templates for  micro/nanoscale devices, and molecular machines. 
Crucially, these microscopic  engines have become a reality on different platforms, including nitrogen-vacancy centers in diamond \cite{JK19}, nuclear magnetic resonance \cite{lutz16,JP29,RJ19,lutz25},
Brownian systems \cite{VB12,IA16}, atomic collisions \cite{QB21}, superconducting circuits \cite{TU26}, and bacteria \cite{SK16}.

However, the work output of a standard microscopic heat engine is  governed by the stochastic trajectory distribution and the  energy spectrum \cite{Sei12,MC11,Lutz07}. When the working system is in a Gibbs thermal state, its trajectory distribution is determined by the reservoir temperatures and energy levels. Consequently, for the engine to operate in the positive-work regime, the permissible range of energy levels is  constrained by the reservoir temperatures \cite{Quan07,TD04}. Furthermore, in accordance with the second law of thermodynamics, for given hot reservoir temperature $T_\mathrm{h}$, the heat absorbed by a $d$-level system from the hot reservoir is bounded by $Q_\mathrm{h}<k_B T_\mathrm{h} \ln d$  with $k_B$ denoting the Boltzmann constant. By the first law of thermodynamics, the extractable work $W< Q_\mathrm{h}$ is thus limited by $T_\mathrm{h}$. As a result, once the reservoir temperatures are fixed, both the operational range of the energy level  (robustness against parameter tuning) and the  work output are constrained.

Another  bottleneck restricting the practical utility of the microscopic heat engines is the thermal fluctuations induced by the heat reservoirs \cite{Sei12,lutz20,Ma22,Wang19}. These fluctuations render the work extracted per cycle a stochastic variable. The resulting work fluctuations  undermine the  reliability of the engine. For instance, when some stochastic trajectories yield zero or negative work, even having a positive mean work output, the engine may fail to deliver useful work  in a considerable fraction of cycles, or even require external work input,  thereby preventing stable power delivery.

%\st{However, these  constraints can be circumvented by employing continuous Maxwell's demon (CMD). The CMD is a feedback control device that continuously monitors the  system state until a predefined target state is detected, at which point feedback operations are triggered, and  has already been demonstrated experimentally. By utilizing CMD, one can enforce identical trajectory during every  cycle, which  not only decouples the work output from reservoir temperatures, making it depend solely on the system's energy level, but also completely eliminates thermal  fluctuations.}
%\textcolor{red}{This paragraph sounds like the basic mechanism has been worked, we only give an example. This destroys the importance of this work. I suggest to shorten it and integrate to the next paragraph, emphasizing why what we do here is important and innovative.}

\begin{figure*}
\centering
    \includegraphics[width=1\linewidth]{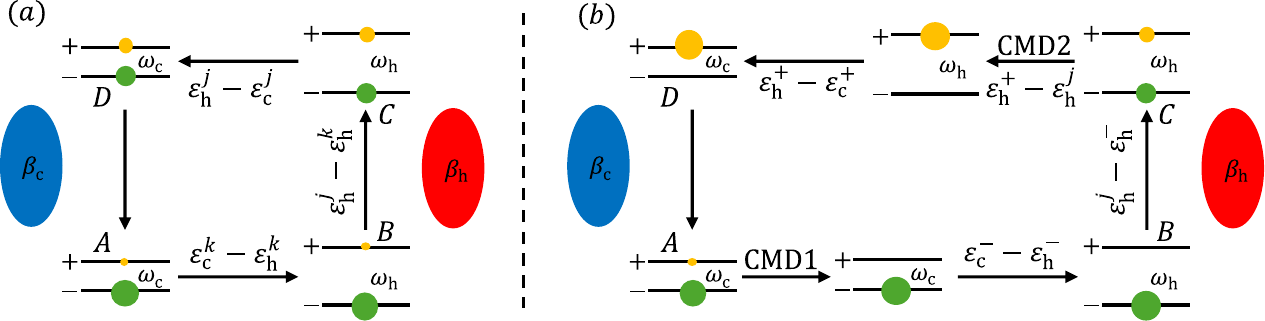}
    \caption{Schematic diagrams of the Otto heat engine (OHE) (a) and the Otto information engine (OIE) (b). The OHE  consists of four strokes: Adiabatic compression $A\to B$, isochoric heating $B\to C$, adiabatic expansion $C\to D$, isochoric cooling $D\to A$. In the OIE, continuous Maxwell's demon 1 (CMD1) and  2 (CMD2) are introduced at points $A$ and $C$ of the  OHE, respectively, ensuring that the adiabatic compression stroke $A \to B$ and the adiabatic expansion stroke $C \to D$ operate deterministically in  ground state and  excited state, respectively. The green  and yellow balls represent the probabilities of the system being in the ground   and  excited  states, respectively.}
    \label{model}
\end{figure*}

{To overcome these limitations, we incorporate two continuous Maxwell's demons (CMDs) \cite{MR19,GM21,TT25,MR21,JP23} into an  Otto heat engine, constructing an  Otto information engine (OIE). The CMD is a feedback control device that continuously monitors the  system state until a predefined target state is detected, at which point feedback operations are triggered, and  has already been demonstrated experimentally \cite{MR21}. Then, we utilize the  demons to achieve that the operational trajectory of the OIE is the  same from cycle to cycle.} Consequently, the output  work of the OIE depends exclusively on the energy level gap, enabling an arbitrary  size work output with zero fluctuations. Furthermore, we demonstrate that even when accounting for the energetic cost required to erase the memories of both CMDs, the OIE's efficiency can still surpass the standard Carnot efficiency. Finally, our finite-time analysis indicates that even  incorporating the measurement time of the CMDs, it can achieve a drastic enhancement in output power compared to its conventional counterpart. To better illustrate this point, we also conducted Monte Carlo simulations using the experimental parameters from Ref. \cite{JP29}.

\section{Limitations of the Otto Heat Engines}
As illustrated in Fig.~\ref{model}(a), a conventional quantum Otto heat engine (OHE) comprises two adiabatic and two isochoric strokes. Here, the working system of the OHE is a spin-$1/2$ system with energy level gap $\omega$. We first consider the quasistatic case. During the two adiabatic strokes, the working system is decoupled from the heat reservoirs, so no heat exchange occurs. The populations of the system remain invariant, and the work done  by the system equals its internal energy change. Consequently, the work performed during the adiabatic compression ($A \to B$) and adiabatic expansion ($C \to D$) strokes are given by $\varepsilon_\mathrm{c}^k-\varepsilon_\mathrm{h}^k$ and $\varepsilon_\mathrm{h}^j-\varepsilon_\mathrm{c}^j$, respectively, where $\varepsilon_\mathrm{c,h}^k=k\omega_\mathrm{c,h}/2$ and $ \varepsilon_\mathrm{h,c}^j=j\omega_\mathrm{h,c}/2$ ($2\pi\hbar=1$) denote the initial and final eigenenergies of these two adiabatic strokes with $k, j \in \{+, -\}$ labeling the states of the system at points $A$ and $C$. Here, $\omega_\mathrm{c}$ and $\omega_\mathrm{h}>\omega_\mathrm{c}$ represent the energy level gaps at points $A$ ($D$) and $C$ ($B$), respectively.

During the cold and hot isochoric strokes, the energy level gaps of the system are fixed at $\omega_\mathrm{c}$ and $\omega_\mathrm{h}$, while the system is  coupled to heat reservoirs at inverse temperatures $\beta_\mathrm{c} = 1/(k_B T_\mathrm{c})$ and $\beta_\mathrm{h} = 1/(k_B T_\mathrm{h})$, respectively. At the end of the two isochoric strokes (points $A$ and $C$), the system relaxes to the respective thermal Gibbs states, with the occupation probabilities of states $|k\rangle$ and $|j\rangle$ given by $p_\mathrm{c}^k = e^{-\beta_\mathrm{c} \varepsilon_\mathrm{c}^k}/(e^{-\beta_\mathrm{c} \varepsilon_\mathrm{c}^+}+e^{-\beta_\mathrm{c} \varepsilon_\mathrm{c}^-})$ and $p_\mathrm{h}^j = e^{-\beta_\mathrm{h} \varepsilon_\mathrm{h}^j}/(e^{-\beta_\mathrm{h} \varepsilon_\mathrm{h}^+}+e^{-\beta_\mathrm{h} \varepsilon_\mathrm{h}^-})$. Since no work is extracted along the isochore stroke, the heat absorbed during the hot isochoric stroke matches the change in internal energy, $q_\mathrm{h}=  \varepsilon_\mathrm{h}^j-\varepsilon_\mathrm{c}^k$, and the total  work over a  cycle reads $w = \varepsilon_\mathrm{c}^k-\varepsilon_\mathrm{h}^k +\varepsilon_\mathrm{h}^j-\varepsilon_\mathrm{c}^j$. Consequently, the probability distributions of the stochastic work output $p(w)=\sum_{h,j} p_c^k p_h^j\delta[w-(\varepsilon_\mathrm{c}^k-\varepsilon_\mathrm{h}^k +\varepsilon_\mathrm{h}^j-\varepsilon_\mathrm{c}^j)]$ and heat $p(q_h )=\sum_{k,j}p_\mathrm{c}^k p_\mathrm{h}^j\delta[q_\mathrm{h}-(\varepsilon_\mathrm{h}^j-\varepsilon_\mathrm{h}^k)]$ input per cycle are given by
\begin{eqnarray}
 p(w)&=&p_\mathrm{c}^+ p_\mathrm{h}^- \delta[w-(\omega_\mathrm{c}-\omega_\mathrm{h} )]+p_\mathrm{c}^- p_\mathrm{h}^+ \delta[w-(\omega_\mathrm{h}-\omega_\mathrm{c})]\nonumber\\
& +&(p_\mathrm{c}^+ p_\mathrm{h}^++p_\mathrm{c}^- p_\mathrm{h}^- )\delta(w),\label{pw}\\
  p(q_\mathrm{h})&=&p_\mathrm{c}^+ p_\mathrm{h}^- \delta(q_\mathrm{h}+\omega_\mathrm{h})+p_\mathrm{c}^- p_\mathrm{h}^+ \delta(q_\mathrm{h}-\omega_\mathrm{h})\nonumber\\
& +&(p_\mathrm{c}^+ p_\mathrm{h}^++p_\mathrm{c}^- p_\mathrm{h}^- )\delta(q_\mathrm{h}).
\end{eqnarray}
Here, $p_\mathrm{c}^kp_\mathrm{h}^j$ represent the probability of the trajectory $(k,j)$.

Then, the average work output $W_\mathrm{OHE}^\mathrm{qs}=\sum_w p(w)w$  and the average heat $Q_\mathrm{OHE}^\mathrm{qs}=\sum_{q_\mathrm{h}} p(q_\mathrm{h})q_\mathrm{h}$ absorbed from the hot reservoir of the OHE are given by
\begin{align}
W_\mathrm{OHE}^\mathrm{qs} &= (\omega_\mathrm{h} - \omega_\mathrm{c})(\langle n_\mathrm{h}\rangle - \langle  n_\mathrm{c}\rangle),\label{WOHEqs}\\
Q_\mathrm{OHE}^\mathrm{qs} &= \omega_\mathrm{h}(\langle n_\mathrm{h}\rangle - \langle  n_\mathrm{c}\rangle),\label{QOHEqs}
\end{align}
where $\langle n_\mathrm{c}\rangle = (p_\mathrm{c}^+-p_\mathrm{c}^-)/2=-\tanh(\beta_\mathrm{c}\omega_\mathrm{c}/2)/2$ and $\langle n_\mathrm{h}\rangle = (p_\mathrm{h}^+-p_\mathrm{h}^-)/2=-\tanh(\beta_\mathrm{h}\omega_\mathrm{h}/2)/2$ represent the average quantum number of the system at points $A$ and $C$, respectively. 
According to the first law of thermodynamics, the average heat exchanged with the cold reservoir is $W_\mathrm{OHE}^\mathrm{qs}-Q_\mathrm{OHE}^\mathrm{qs}$. 

Eq. (\ref{WOHEqs}) indicates that the regime for the OHE produces positive work is    $\langle n_\mathrm{h}\rangle > \langle n_\mathrm{c}\rangle$, leading to $\omega_\mathrm{c}<\omega_\mathrm{h}< \omega_\mathrm{c}\beta_\mathrm{c}/\beta_\mathrm{h}$. Furthermore, for  fixed reservoir temperatures, the average quantum numbers $\langle n_\mathrm{c,h}\rangle$ monotonically decrease as the energy level gap $\omega_\mathrm{c,h}$ increases. In the  limits  $\omega_\mathrm{c} \gg 1$ and $\omega_\mathrm{c} \to 0$, where $\omega_\mathrm{h} \gg 1$ and $\omega_\mathrm{h} \to 0$, these average quantum numbers satisfy $\langle n_\mathrm{c,h}\rangle\to-0.5$ and $\langle n_\mathrm{c,h}\rangle\to0$,
leading    $\langle n_\mathrm{h}\rangle \approx \langle n_\mathrm{c}\rangle$. Therefore, for fixed thermal reservoirs temperatures, the energy level gaps $\omega_\mathrm{c}$ and $\omega_\mathrm{h}$ cannot be chosen arbitrarily. Their operational parameter window is strictly bounded by how the extractable work approaches zero, which compromises the robustness of the OHE against energy level gap. 

Furthermore, according to Eq.~(\ref{QOHEqs}), the maximum heat absorbed from the hot reservoir is bounded by the hot reservoir temperature. The  derivation is as follows. In the positive temperature regime, since $-1/2<\langle n_\mathrm{c}\rangle < 0$, we have  $Q_\mathrm{OHE}^\mathrm{qs} <   f(\lambda)=\lambda[1-\tanh(\lambda)]/\beta_\mathrm{h}$, where $\lambda = \beta_\mathrm{h} \omega_\mathrm{h}/2>0$. Differentiating $f(\lambda)$ with respect to $\lambda$ yields $[1-\tanh(\lambda)]\{1-\lambda[1+\tanh(\lambda)]\} $. Setting $f'(\lambda) = 0$ leads to 
$1-\lambda[1+\tanh(\lambda)]=0$.
The unique  root of this equation is $\lambda^* \approx 0.64$, at which $f(\lambda)$ attains its  maximum $f(\lambda^*) \approx 0.28$. Consequently, the  heat input  is bounded by $Q_\mathrm{OHE}^\mathrm{qs} < 0.28/\beta_\mathrm{h}<\ln 2/\beta_\mathrm{h}$. Since the extracted work strictly satisfies $W_\mathrm{OHE}^\mathrm{qs}  < Q_\mathrm{OHE}^\mathrm{qs}$, the  work output of the OHE is inherently capped once the hot reservoir temperature is fixed.

On the other hand, according to the distribution of the work Eq. (\ref{pw}), different operating trajectories will generate different output work. Specifically,  along trajectories  $(k=-, j=+)$, $(k=+, j=-)$, $(k=+, j=+)$, and $(k=-, j=-)$, the output work of the OHE  are  $\omega_\mathrm{h} - \omega_\mathrm{c}$, $-(\omega_\mathrm{h} - \omega_\mathrm{c})$, $0$, and $0$, with corresponding probabilities $p_\mathrm{c}^- p_\mathrm{h}^+,p_\mathrm{c}^+ p_\mathrm{h}^-,p_\mathrm{c}^+ p_\mathrm{h}^+$, and $p_\mathrm{c}^- p_\mathrm{h}^-$, respectively. Consequently, the work extracted from the OHE is  fluctuating. Such fucutuations hinder the practical performance of the OHE. For example, when the OHE is coupled to downstream nanoscale quantum batteries or microscopic molecular loads, large work fluctuations can  degrade energy storage fidelity or even trigger stochastic engine reversal due to negative-work realization events. Therefore, suppressing or completely eliminating work fluctuations is essential. According to  Eq. (\ref{pw}), the fluctuations $\delta w_\mathrm{OHE}^\mathrm{qs^{2}}=\langle w^2\rangle-(W_\mathrm{OHE}^\mathrm{qs})^2$ of the work  can be expressed as
\begin{equation}\label{dwqs}
\delta w_\mathrm{OHE}^\mathrm{qs^{2}} =  (\omega_\mathrm{h} - \omega_\mathrm{c})^2 (\frac{1}{2}-\langle n_\mathrm{h}\rangle^2-\langle n_\mathrm{c}\rangle^2).
\end{equation}

Finally, the thermodynamic efficiency of the heat engine is defined as the ratio of the average work output to the average heat absorbed from the hot reservoir. Utilizing Eqs. (\ref{WOHEqs}) and (\ref{QOHEqs}), the efficiency of the OHE reads
\begin{equation}
\eta_\mathrm{OHE}^\mathrm{qs} =\frac{W_\mathrm{OHE}^\mathrm{qs}}{Q_\mathrm{OHE}^\mathrm{qs}}  = 1 - \frac{\omega_\mathrm{c}}{\omega_\mathrm{h}}.
\end{equation}
Under positive work condition $\omega_\mathrm{h}<\omega_\mathrm{c}\beta_\mathrm{c}/\beta_\mathrm{h}$, the efficiency of the OHE is strictly bounded by the standard Carnot efficiency, i.e.,
$\eta_\mathrm{OHE}^\mathrm{qs}  < \eta_{\rm C}=1 - \beta_\mathrm{h}/\beta_\mathrm{c}$,
where $\eta_{\rm C}$ denotes the Carnot efficiency.

\section{Otto Information Engine}

In summary, the limited robustness against energy level gap variations, the  bounded work output, the stochastic work fluctuations, and the Carnot-bounded efficiency in the OHE all originate from the intrinsic dependence of the stochastic trajectory distribution on the energy level gaps. To overcome these limitations, it is essential to decouple the system's trajectory distribution from its energy level gaps. The recently proposed continuous Maxwell's demon (CMD) offers a viable mechanism to achieve this goal. A CMD operates by continuously monitoring the  system state and applying immediate feedback control as soon as a prescribed transition in the system's state is detected \cite{MR19}.

Because the trajectory distribution in the OHE is governed by the thermal Gibbs distributions $p_{\mathrm{c}}^k$ and $p_{\mathrm{h}}^j$, %(or equivalently, the average quantum numbers $\langle n_{\mathrm{c}} \rangle$ and $\langle n_{\mathrm{h}} \rangle$), 
 we incorporate two error-free CMDs into the OHE, called  Otto information engine (OIE), as illustrated in Fig.~\ref{model}(b). The specific intervention stages of the two demons are implemented as follows.

At point $A$ ,  the first continuous Maxwell's demon (CMD1) is introduced to continuously monitor the system  until the ground state $|-\rangle$ is detected, immediately upon which the adiabatic compression stroke ($A \to B$) is triggered. As a consequence, the initial state of the adiabatic compression stroke %in every  cycle of 
the OIE is  initialized in the ground state. Subsequently, the second continuous Maxwell's demon (CMD2) is deployed at point $C$ to continuously monitor the system's state until the excited state $|+\rangle$ is detected, whereupon the adiabatic expansion stroke ($C \to D$) is initiated. Thus, the initial state of the adiabatic expansion stroke %in every cycle 
of the OIE
is deterministically locked into the excited state. Through this dual-demon protocol, the system deterministically follows the trajectory $(k=-, j=+)$ in every cycle,   decoupling the trajectory distribution  from the energy level gaps and reservoir temperatures.

In the OIE, the average  quantum numbers at the initial moments of the adiabatic compression and expansion strokes of the system are
$\langle n_{\mathrm{c}}^\mathrm{OIE} \rangle = -1/2$ and $\langle n_{\mathrm{h}}^\mathrm{OIE} \rangle = 1/2$, respectively. Replacing $\langle n_{\mathrm{c,h}}\rangle $ with $\langle n_{\mathrm{c,h}}^\mathrm{OIE} \rangle$ in Eqs. (\ref{WOHEqs})-(\ref{dwqs}), the work output work,  the total energy injected, and the corresponding work fluctuations of the OIE are obtained as
\begin{align}
W_{\mathrm{OIE}}^\mathrm{qs} &=\omega_{\mathrm{h}} - \omega_{\mathrm{c}},\label{WOIEqs}\\
E_{\mathrm{OIE}}^\mathrm{qs} & = \omega_{\mathrm{h}},\label{QOIEqs}\\
\delta w_\mathrm{OIE}^\mathrm{qs^{2}}&=0.\label{dWOIEqs}
\end{align}
According to Eqs.~(\ref{WOIEqs}) and (\ref{dWOIEqs}), the OIE can produce positive work for any arbitrary energy level gaps satisfying $\omega_{\mathrm{h}} > \omega_{\mathrm{c}}$. Crucially, this work output possesses no thermodynamic upper bound and is completely devoid of thermal fluctuations.

Furthermore, combining Eqs.~(\ref{WOHEqs}) and (\ref{WOIEqs}), for identical energy level gaps, the work produced by the OIE is amplified by a factor of $(\langle n_{\mathrm{h}} \rangle - \langle n_{\mathrm{c}} \rangle)^{-1}\geq2$ compared to that of the OHE. Equality holds only in the low temperature limit $\beta_\mathrm{c}\to \infty$  where $\langle n_{\mathrm{c}} \rangle \to -1/2$ and the high temperature limit $\beta_\mathrm{h}\to 0$  where $\langle n_{\mathrm{h}} \rangle \to 0$. Utilizing  experimental parameters from Ref.~\cite{JP29}, compared to OHE, the OIE's work output can be increased by 28 times.

It is worth emphasizing that in Eq.~(\ref{QOIEqs}), the total injected energy $E_{\mathrm{OIE}}^\mathrm{qs}$  comprises two  contributions: the heat $\omega_\mathrm{h}(\langle n_\mathrm{h}\rangle+1/2)$ absorbed from the hot reservoir and the measurement energy $\omega_\mathrm{h}(1/2-\langle n_\mathrm{h}\rangle)$ induced by the CMD2 via measurement and  feedback control.

To maintain the stable operation of OIE, the  information stored in the memories of both CMDs must be  reset. {Otherwise, the memory acquired in previous cycles would mix with that of the current cycle, preventing the demon's feedback control from perfectly matching the present measurement outcomes}. In accordance with Landauer's principle, the minimal energy  required to erase information $I$ is given by $I/\beta$ \cite{Landauer61,TS09}. To minimize the memory erasure cost without introducing an external auxiliary reservoir, we utilize the  cold reservoir to reset the CMDs's memories. Assuming that the information acquired by CMD1 and CMD2  are $I_{\mathrm{c}}$ and $I_{\mathrm{h}}$, respectively, the total energetic cost required to erase the memories of both demons reads $(I_{\mathrm{c}}+I_{\mathrm{h}})/\beta_\mathrm{c}$. 
Consequently, the total energy injected  into the combined setup (the system and the two CMDs)  is $E_{\mathrm{OIE}}^\mathrm{qs} +  (I_{\mathrm{c}}+I_{\mathrm{h}})/\beta_\mathrm{c}$. Accordingly, the  efficiency $\eta_{\mathrm{OIE}}^\mathrm{qs}={W_{\mathrm{OIE}}^\mathrm{qs}}/[{E_{\mathrm{OIE}} +  (I_{\mathrm{c}}+I_{\mathrm{h}})/\beta_\mathrm{c}}]$ of the OIE is formulated as
\begin{equation}\label{etaOIE}
\eta_{\mathrm{OIE}}^\mathrm{qs} = \frac{\eta_\mathrm{OHE}^\mathrm{qs}}{1+(I_{\mathrm{c}}+I_{\mathrm{h}})/(\beta_\mathrm{c}E_{\mathrm{OIE}})}<\eta_\mathrm{OHE}^\mathrm{qs}.
\end{equation}

In Appendix \ref{app1}, $I_{\mathrm{c}}$ and $I_{\mathrm{h}}$ have determined, which read $I_{\mathrm{c}}=I_\mathrm{min}^\mathrm{c}+I_\tau^\mathrm{c}$, where $I_\mathrm{min}^\mathrm{c} = -p_\mathrm{c}^+ \ln p_\mathrm{c}^+ - p_\mathrm{c}^- \ln p_\mathrm{c}^- - p_\mathrm{c}^+ {p_\mathrm{c}^+}/{p_c^-} \ln p_c^+ - p_\mathrm{c}^+ \ln p_\mathrm{c}^- = -\ln p_\mathrm{c}^- - {p_\mathrm{c}^+}/{p_\mathrm{c}^-}\ln p_\mathrm{c}^+$ and $I_\tau^\mathrm{c} = -p_\mathrm{c}^+ {e^{-\Gamma_\mathrm{c} \tau}}/[p_\mathrm{c}^-(1-e^{-\Gamma_\mathrm{c} \tau})] \ln p_\mathrm{c}^+ - p_\mathrm{c}^+ {(p_\mathrm{c}^+ + p_\mathrm{c}^- e^{-\Gamma_\mathrm{c} \tau})}\ln ( 1 + {p_\mathrm{c}^- e^{-\Gamma_\mathrm{c} \tau}}/{p_\mathrm{c}^+} )/[p_\mathrm{c}^-(1-e^{-\Gamma_\mathrm{c} \tau})]  - p_\mathrm{c}^+ \ln(1 - e^{-\Gamma_\mathrm{c} \tau})$, and   $I_{\mathrm{h}}=I_\mathrm{min}^\mathrm{h}+I_\tau^\mathrm{h}$ where $I_\mathrm{min}^\mathrm{h}=-\ln p_\mathrm{h}^+-p_\mathrm{h}^-/p_\mathrm{h}^+\ln p_\mathrm{h}^-$, and $I_\tau^\mathrm{h}= -p_\mathrm{h}^- {e^{-\Gamma_\mathrm{h} \tau}}/{[p_\mathrm{h}^+(1-e^{-\Gamma_\mathrm{h} \tau})]} \ln p_\mathrm{h}^- - p_\mathrm{h}^- {[p_\mathrm{h}^- + p_\mathrm{h}^+ e^{-\Gamma_\mathrm{h} \tau}]}\ln (1 + {p_\mathrm{h}^+ e^{-\Gamma_\mathrm{h} \tau}}/{p_\mathrm{h}^-} )/{[p_\mathrm{h}^+(1-e^{-\Gamma_\mathrm{h} \tau})]}  - p_\mathrm{h}^- \ln(1 - e^{-\Gamma_\mathrm{h} \tau})$. 
Here,  $\tau$  represents the time interval between two adjacent measurements. The terms  $\Gamma_{\mathrm{c,h}} \equiv \gamma_0 (2N_{\mathrm{c,h}} + 1)$ being the total transition rates, where $N_{\mathrm{c,h}} = (e^{\beta_{\mathrm{c,h}} \omega_{\mathrm{c,h}}} - 1)^{-1}$ are the mean photon number of the cold and hot reservoirs and we have assumed identical spontaneous emission rates $\gamma_0$ for both reservoirs.

Differentiating $I_\tau^\mathrm{c}$ with respect to $x_\mathrm{c}=e^{-\Gamma_{\mathrm{c}}\tau}$ yields
${d I_\tau^\mathrm{c}}/{d x_\mathrm{c}} = -{p_\mathrm{c}^+ \ln(p_\mathrm{c}^+ + p_\mathrm{c}^- x_\mathrm{c})}/{[p_\mathrm{c}^- (x_\mathrm{c}-1)^2]} > 0$.
Since $x_\mathrm{c}$ monotonically decreases as the measurement interval $\tau$ increases, $I^{\mathrm{c}}_\tau$ is a  monotonically decreasing function of $\tau$. In the  limit $\tau \gg 1$, $I^{\mathrm{c}}_\tau$ attains its  minimum $I^{\mathrm{c}}_\tau=0$. Conversely, in the continuous monitoring limit $\tau \to 0$, $I_{\mathrm{c}}$ diverges logarithmically. Therefore, $I_\mathrm{min}^\mathrm{c}$ represents the minimum information acquired by the CMD1. Similarly, $I_\mathrm{min}^\mathrm{h}$is the minimum information obtained by the CMD2.

As a result, 
%Therefore, when no memory correlation is retained between successive measurements performed by the demons,i.e. $\Gamma_\mathrm{h}\tau,\Gamma_\mathrm{c}\tau\gg1$, both CMDs acquire the minimal amount of information: $I_\mathrm{c}=I^{\mathrm{c}}_{\min}$ and $I_\mathrm{c}=I^{\mathrm{h}}_{\min}$. 
in the regime $\Gamma_\mathrm{h}\tau,\Gamma_\mathrm{c}\tau\gg1$, the overall thermodynamic efficiency Eq. (\ref{etaOIE}) of the OIE can be expressed as
$\eta_{\mathrm{OIE}}^\mathrm{qs} = {\eta_\mathrm{OHE}^\mathrm{qs}}/{[1+(I^{\mathrm{c}}_{\min}+I^{\mathrm{h}}_{\min})/(\beta_\mathrm{c}E_{\mathrm{OIE}})]}.$
Furthermore, in the regime $\omega_\mathrm{h}\gg\omega_\mathrm{c}$ where $\eta_\mathrm{OHE}^\mathrm{qs}\to1$ and $E_\mathrm{OIE}\gg I_\mathrm{c}$, 
$\eta_{\mathrm{OIE}}^\mathrm{qs}$ can be  simplified to
$\eta_{\mathrm{OIE}}^\mathrm{qs} = 1/{[1+I^{\mathrm{h}}_{\min}/(\beta_\mathrm{c}\omega_\mathrm{h})]}.$ 
Differentiating $I^{\mathrm{h}}_{\min}/(\beta_\mathrm{c}\omega_\mathrm{h})$ with respect to $\omega_{\mathrm{h}}$ yields
$[d I^{\mathrm{h}}_{\min}/(\beta_\mathrm{c} \omega_\mathrm{h})]/{d \omega_\mathrm{h}} = \{-\ln(1 + e^{\beta_\mathrm{h} \omega_\mathrm{h}}) - e^{\beta_\mathrm{h} \omega_\mathrm{h}}(\beta_\mathrm{h} \omega_\mathrm{h} - 1)[\beta_\mathrm{h} \omega_\mathrm{h} - \ln(1 + e^{\beta_\mathrm{h} \omega_\mathrm{h}})]\}/(\beta_\mathrm{c} {\omega_\mathrm{h}}^2)$.
Given $\omega_{\mathrm{h}} \gg \omega_{\mathrm{c}}$ where $\ln(1 + e^{\beta_\mathrm{h}\omega_\mathrm{h}})\approx \beta_\mathrm{h}\omega_\mathrm{h}$, the derivative remains  negative, confirming that $\eta_{\mathrm{OIE}}$ monotonically increases with $\omega_{\mathrm{h}}$.
Because  $\lim_{\omega_{\mathrm{h}} \to \infty} I^{\mathrm{h}}_{\min}/(\beta_\mathrm{c} \omega_\mathrm{h}) = 1-\eta_\mathrm{C}$ where $\eta_{\mathrm{OIE}}^\mathrm{qs}=1/(2-\eta_\mathrm{C})> \eta_{\mathrm{C}}$, there always exists a critical $\omega_{\mathrm{h}}$  for   $\eta_{\mathrm{OIE}}^\mathrm{qs} > \eta_{\mathrm{C}}$. 
Consequently, the efficiency of the OIE can  surpass the  Carnot efficiency. Crucially, this mechanism enables the OIE to operate when the two reservoir temperatures are identical $\eta_{\mathrm{C}}=0$,  where the OHE ceases to operate. 

In summary, the integration of the CMD enhances the robustness of the quantum Otto engine against energy level gap variations, removes the thermodynamic bound on work extraction, completely eliminates thermal work fluctuations, and enables super-Carnot operational efficiency.

\section{Finite time case}

Although the continuous measurements executed by the CMDs introduce an additional operational time, %$\tau_{\mathrm{M}}^{\mathrm{c}} + \tau_{\mathrm{M}}^{\mathrm{h}}$, 
the introduction of the CMDs  brings about a substantial multiplication in the output power, as elucidated below.

During the adiabatic compression and expansion strokes, the time-dependent Hamiltonians of the system are respectively parameterized as
\begin{align}
H_{\mathrm{com}}(t) &= \omega_{\mathrm{ch}}(t) \big\{ [\cos\theta(t)] \sigma_x + [\sin\theta(t)] \sigma_z \big\},\label{Hcom} \\
H_{\mathrm{exp}}(t) &= H_{\mathrm{com}}(\tau_{\mathrm{dri}} - t),
\end{align}
where $\omega_{\mathrm{ch}}(t) = \frac{\omega_{\mathrm{c}}}{2}(1 - t/\tau_{\mathrm{dri}}) + \frac{\omega_{\mathrm{h}}}{2} t/\tau_{\mathrm{dri}}$, and the driving angle  $\theta(t) = \frac{\pi}{2} [3(t/\tau_{\mathrm{dri}})^2 - 2(t/\tau_{\mathrm{dri}})^3]$ with $\tau_{\mathrm{dri}}$ denoting the driving time. 
Here, we have assumed that the adiabatic expansion stroke is the  time-reversed counterpart of the adiabatic compression stroke.  {The reason we use this Hamiltonian is that its similar form has already been experimentally implemented \cite{JP29}.}

\begin{figure}
    \centering
    \includegraphics[width=0.8\linewidth]{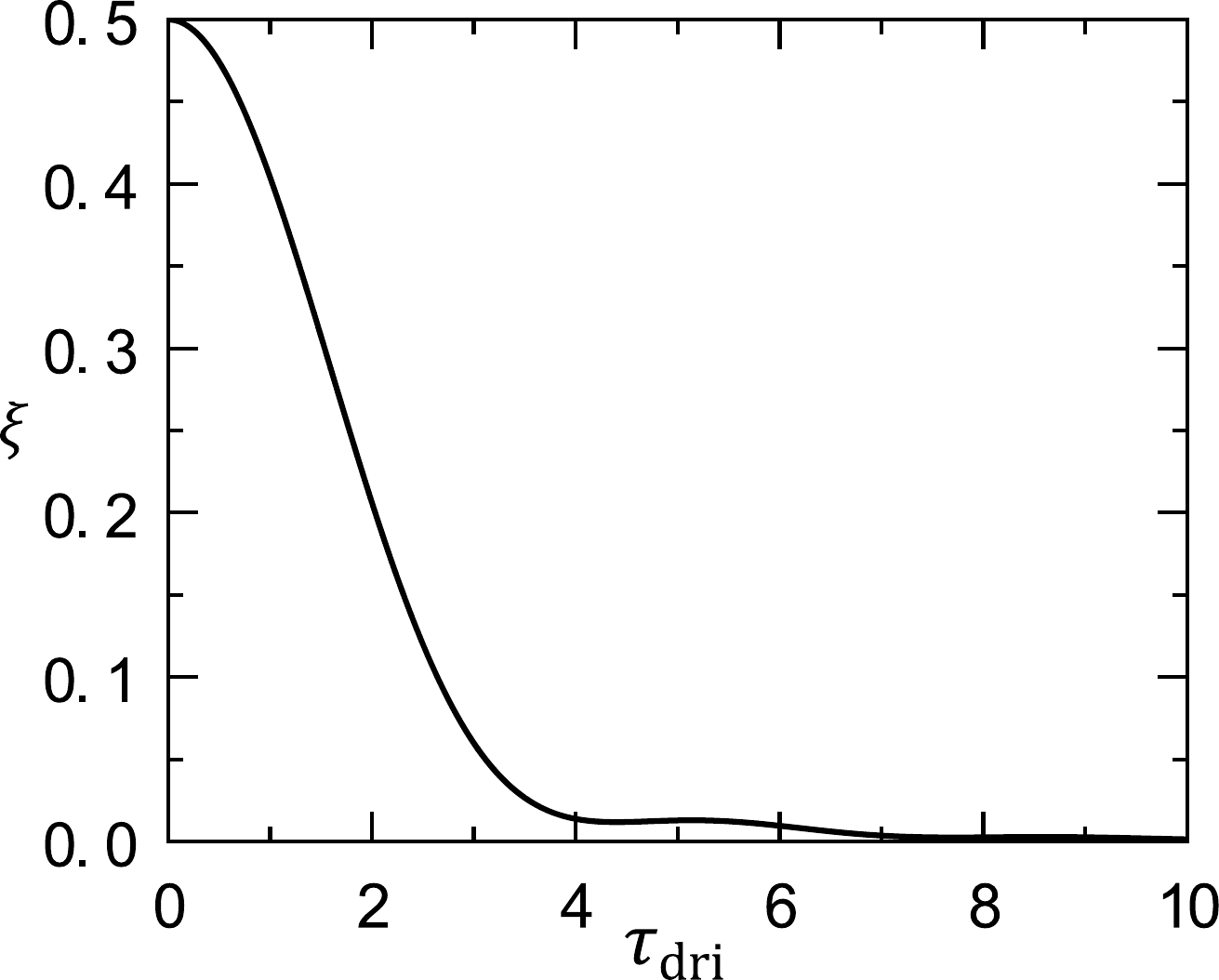}
    \caption{The transition probability $\xi$ as  a function of the driving time $\tau_\mathrm{dri}$. The parameters are $\beta_\mathrm{c}=1$, $\beta_\mathrm{h}=0.3$, $\omega_\mathrm{c}=1$, and $\omega_\mathrm{h}=3$.}
    \label{xi}
\end{figure}

Because the Hamiltonians at different time do not commute, i.e., $[H(t), H(t')] \neq 0$,  there are nonadiabatic transitions between distinct energy levels. The corresponding transition probability is defined as
$ \xi= |\langle j \neq k | U_{\mathrm{ch}} | k \rangle|^2 = |\langle k \neq j | U_{\mathrm{hc}} | j \rangle|^2
$ \cite{JP29}
where $U_{\mathrm{ch}} = \mathcal{T} \exp\big(-{i} \int_0^{\tau_{\mathrm{dri}}} H_{\mathrm{com}}(t)\,\mathrm{d}t\big)$ represents the time-evolution operator along the compression stroke, with $\mathcal{T}$ being the time-ordering operator, and $U_{\mathrm{hc}}$ corresponds to the expansion stroke. %Here,  $|k\rangle$ and $|j\rangle$ represent the instantaneous eigenstates of $H_{\mathrm{com}}(0)$ and $H_{\mathrm{com}}(\tau_{\mathrm{dri}})$, respectively
Correspondingly,  $1 - \xi = |\langle j = k | U_{\mathrm{ch}} | k \rangle|^2 = |\langle k = j | U_{\mathrm{hc}} | j \rangle|^2$ is the probability for the system to remain in the same state during the unitary strokes.  In the quasistatic driving limit ($\tau_{\mathrm{dri}} \to \infty$), the system's density matrix commutes with the instantaneous Hamiltonian, precluding nonadiabatic transitions such that $\lim_{\tau_{\mathrm{dri}} \to \infty} \xi = 0$. Conversely, in the sudden-quench limit ($\tau_{\mathrm{dri}} \to 0$), the evolution operator approaches the identity operator, leading to $\xi = |\langle j \neq k | k \rangle|^2 = |\langle k \neq j | j \rangle|^2 = 0.5$. Thus, the nonadiabatic transition probability is  bounded within $0 \le \xi \le 0.5$ and increases as the driving time $\tau_{\mathrm{dri}}$ decreases, as shown in Fig. \ref{xi}.

Employing the two-point measurement scheme \cite{Lutz07}, the stochastic work  generated during the adiabatic compression and expansion strokes are given by $\varepsilon_{\mathrm{c}}^{k} - \varepsilon_{\mathrm{h}}^{k'}$ and $\varepsilon_{\mathrm{h}}^{j} - \varepsilon_{\mathrm{c}}^{j'}$, respectively, where $k', j' \in \{+, -\}$ label the eigenstates of the Hamiltonian at points $B$ and $D$. The heat absorbed during the hot isochoric stroke is $q_{\mathrm{h}}^\mathrm{fin} = \varepsilon_{\mathrm{h}}^{j} - \varepsilon_{\mathrm{h}}^{k'}$. Because the projective measurement destroy quantum coherence, we assume that the working system relaxes  to the thermal Gibbs state at the end of each isochoric stroke, thereby completing dephasing. Specifically, we set the relaxation durations in the hot and cold isochorics strokes  as $\tau_{\mathrm{h}} = 20/\Gamma_{\mathrm{h}}$ and $\tau_{\mathrm{c}} = 20/\Gamma_{\mathrm{c}} > \tau_{\mathrm{h}}$, respectively, ensuring that $\exp(-\Gamma_{\mathrm{h},\mathrm{c}}\tau_{\mathrm{h},\mathrm{c}}) \approx 2 \times 10^{-9}$. In accordance with Eq. (\ref{sol}), this condition guarantees that the system  equilibrates to the respective thermal Gibbs states. Therefore, the occupation probabilities of states $|k\rangle$ and $|j\rangle$ at the beginning of the compression and expansion strokes remain $p_{\mathrm{c}}^{k}$ and $p_{\mathrm{h}}^{j}$. Consequently, the probability distributions of the stochastic work output $w^\mathrm{fin}=\varepsilon_{\mathrm{c}}^{k} - \varepsilon_{\mathrm{h}}^{k'}+\varepsilon_{\mathrm{c}}^{j} - \varepsilon_{\mathrm{h}}^{j'}$ and heat $q_{\mathrm{h}}^\mathrm{fin}$ absorbed from the hot reservoir of such a finite-time  OHE are formulated as
\begin{align}
p(w^\mathrm{fin}) &= \sum_{k,k',j,j'} p_{\mathrm{c}}^{k} |\langle k' | U_{\mathrm{ch}} | k \rangle|^2 p_{\mathrm{h}}^{j} |\langle j' | U_{\mathrm{hc}} | j \rangle|^2\nonumber\\ 
&\times \delta[w^\mathrm{fin} - \big[(\varepsilon_{\mathrm{c}}^{k} - \varepsilon_{\mathrm{h}}^{k'}) + (\varepsilon_{\mathrm{h}}^{j} - \varepsilon_{\mathrm{c}}^{j'})\big]],\label{pwfin}\\
P(q_{\mathrm{h}}^\mathrm{fin}) &= \sum_{k,k',j}  p_{\mathrm{c}}^{k} |\langle k' | U_{\mathrm{ch}} | k \rangle|^2 p_{\mathrm{h}}^{j} \, \delta[q_{\mathrm{h}}^\mathrm{fin} - (\varepsilon_{\mathrm{h}}^{j} - \varepsilon_{\mathrm{h}}^{k'})].
\end{align}

Then, the average work output $W_\mathrm{OHE}^\mathrm{fin} = \sum_{w^\mathrm{fin}} w^\mathrm{fin} p(w^\mathrm{fin})$ and the average heat $Q_{\mathrm{OHE}}^\mathrm{fin} = \sum_{q_{\mathrm{h}}^\mathrm{fin}} q_{\mathrm{h}}^\mathrm{fin} p(q_{\mathrm{h}}^\mathrm{fin})$ absorbed from the hot reservoir of the finite-time OHE can be derived 
\begin{align}
W_\mathrm{OHE}^\mathrm{fin}& = \omega_{\mathrm{c}}\langle n_{\mathrm{c}} \rangle - \omega_{\mathrm{h}}(1 - 2\xi)\langle n_{\mathrm{c}} \rangle \nonumber \\
&+ \omega_{\mathrm{h}}\langle n_{\mathrm{h}} \rangle - \omega_{\mathrm{c}}(1 - 2\xi)\langle n_{\mathrm{h}} \rangle,\label{wOHEfin}\\
Q_{\mathrm{OHE}}^\mathrm{fin}& = \omega_{\mathrm{h}}\big[\langle n_{\mathrm{h}} \rangle - (1 - 2\xi)\langle n_{\mathrm{c}} \rangle\big],\label{qOHEfin}
\end{align}
where $(1 - 2\xi)\langle n_{\mathrm{c}} \rangle$ and $(1 - 2\xi)\langle n_{\mathrm{h}} \rangle$ represent the average quantum numbers at the end of the adiabatic compression and expansion strokes, respectively. From Eq.~(\ref{wOHEfin}), the  condition for the system to extract positive  work is
\begin{equation}
\xi < \xi_0=\frac{\eta_\mathrm{OHE}^\mathrm{qs}}{2} + \frac{1}{2} \frac{(\omega_{\mathrm{c}}^2 - \omega_{\mathrm{h}}^2)\langle n_{\mathrm{h}} \rangle}{\omega_{\mathrm{h}}\langle n_{\mathrm{c}} \rangle + \omega_{\mathrm{c}}\langle n_{\mathrm{h}} \rangle} < \frac{\eta_{\mathrm{O}}}{2} < \frac{1}{2}.
\end{equation}
 Supposing that when $\tau_{\mathrm{dri}} = \tau_0$, $\xi = \xi_0$  where $W_\mathrm{OHE}^\mathrm{fin}=0$, the  OHE cannot extract positive work in the regime $0 < \tau_{\mathrm{dri}} < \tau_0$,  since that  $\xi$ increases  as $\tau_{\mathrm{dri}}$ decreases. Finally, based on Eq. (\ref{wOHEfin}),
the  output power of the finite-time  OHE can be obtained 
$P_\mathrm{OHE} = {W_\mathrm{OHE}^\mathrm{fin}}/{(2\tau_{\mathrm{dri}} + \tau_{\mathrm{h}} + \tau_{\mathrm{c}})}$.

Utilizing  Eq. (\ref{pwfin}), the work fluctuations $\delta w_\mathrm{OHE}^\mathrm{fin^{2}}=\langle w^{\mathrm{fin}^2}\rangle-(W_\mathrm{OHE}^\mathrm{fin})^2$  of the finite-time OHE can be  determined as
\begin{align}\label{dwOHEfin}
\delta w_\mathrm{OHE}^\mathrm{fin^{2}} &= \delta w_\mathrm{OHE}^\mathrm{qs^{2}} + [4\langle n_{\mathrm{h}} \rangle^2 \omega_{\mathrm{c}}^2 + 2(1 - 2\langle n_{\mathrm{h}} \rangle^2 - 2\langle n_{\mathrm{c}} \rangle^2)\nonumber\\
&\times\omega_{\mathrm{c}}\omega_{\mathrm{h}} + 4\langle n_{\mathrm{c}} \rangle^2 \omega_{\mathrm{h}}^2]\xi - 4(\langle n_{\mathrm{h}} \rangle^2 \omega_{\mathrm{c}}^2 + \langle n_{\mathrm{c}} \rangle^2 \omega_{\mathrm{h}}^2)\xi^2.
\end{align}

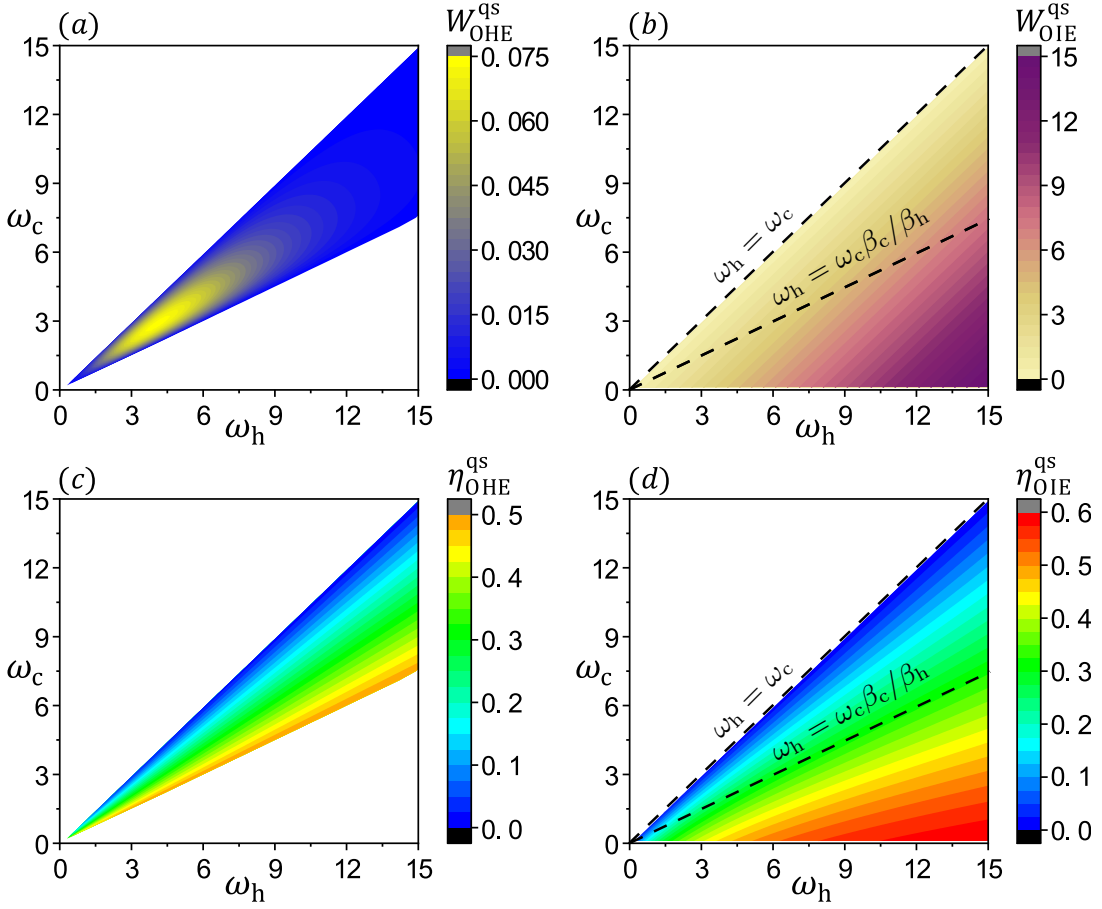
\begin{figure*}
\begin{overpic}[width=0.8\linewidth]{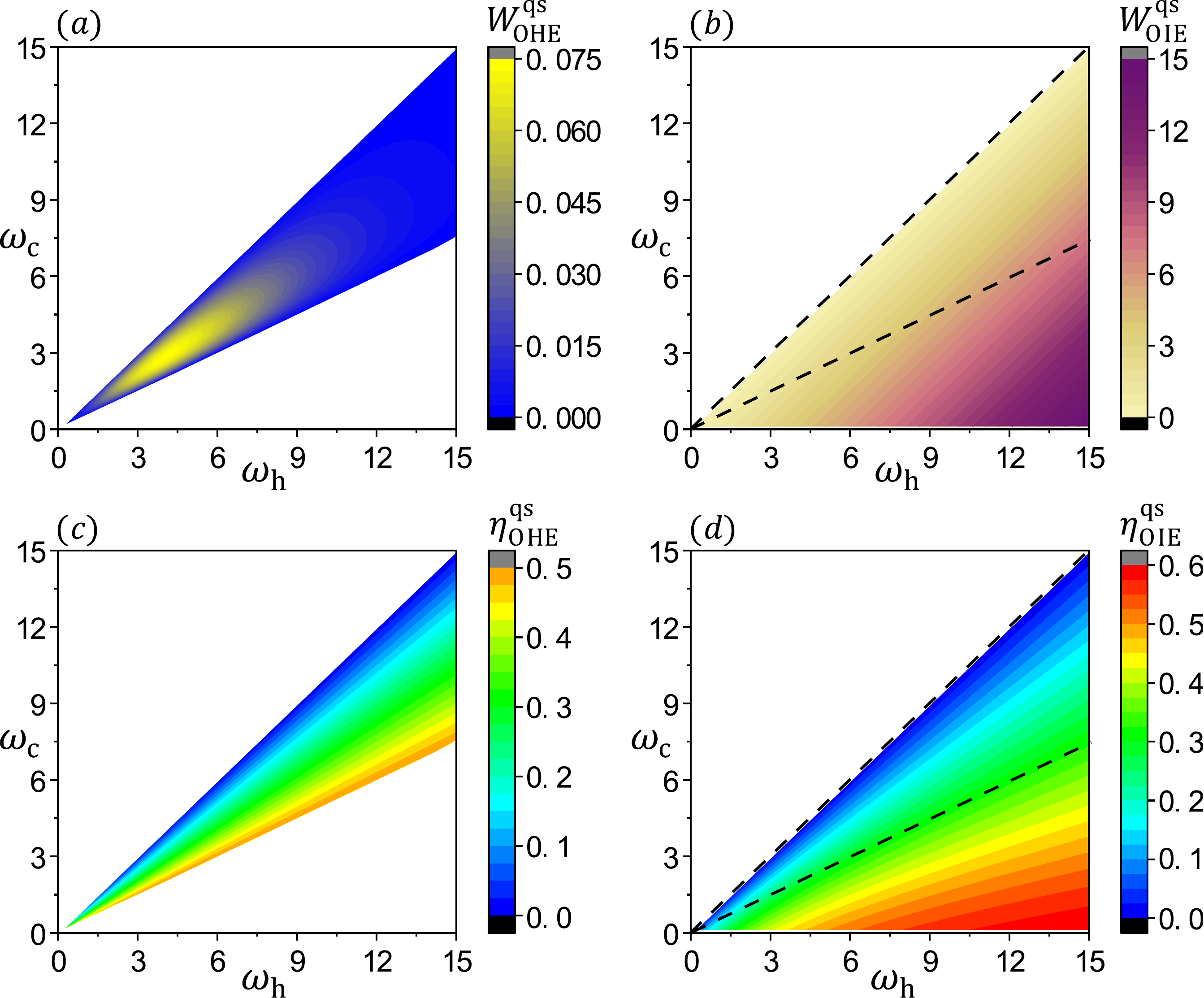}
\put(65,15){\scalebox{1.2}{\rotatebox{45}{$\omega_\mathrm{h}=\omega_\mathrm{c}$}}}
\put(65,57){\scalebox{1.2}{\rotatebox{45}{$\omega_\mathrm{h}=\omega_\mathrm{c}$}}}
\put(70,13){\scalebox{1.2}{\rotatebox{30}{$\omega_\mathrm{h}=\omega_\mathrm{c}\beta_\mathrm{c}/\beta_\mathrm{h}$}}}
\put(70,55){\scalebox{1.2}{\rotatebox{30}{$\omega_\mathrm{h}=\omega_\mathrm{c}\beta_\mathrm{c}/\beta_\mathrm{h}$}}}
\end{overpic}
\caption{The output work and efficiency of the OHE and the OIE as functions of the energy gap $\omega_\mathrm{c}$ and $\omega_\mathrm{h}$. The parameters are $\beta_\mathrm{c}=1$ and $\beta_\mathrm{h}=0.5$ where $\eta_\mathrm{C}=0.5.$}
    \label{fig}
\end{figure*}

 Finally, from Eqs. (\ref{wOHEfin}) and (\ref{qOHEfin}), the  efficiency of the finite-time OHE reads
\begin{equation}\label{effOHEfin}
\eta_\mathrm{OHE}^{\mathrm{fin}} = \frac{W_\mathrm{OHE}^\mathrm{fin}}{Q_{\mathrm{OHE}}^\mathrm{fin}} = \eta_\mathrm{OHE}^\mathrm{qs} + \frac{2\omega_{\mathrm{c}}\xi(\langle n_{\mathrm{c}} \rangle + \langle n_{\mathrm{h}} \rangle)}{Q_{\mathrm{OHE}}^\mathrm{fin}}.
\end{equation}

Similar to the quasistatic case, the finite-time OIE operates by introducing CMD1 and CMD2 at the finite-time OHE's points $A$ and point $C$, respectively.
Substituting the average quantum numbers $\langle n_{\mathrm{c}}^{\mathrm{OIE}} \rangle = -1/2$ and $\langle n_{\mathrm{h}}^{\mathrm{OIE}} \rangle = 1/2$ into Eqs. (\ref{wOHEfin}), (\ref{qOHEfin}), and (\ref{dwOHEfin}), the  output work of the finite-time OIE, the corresponding injected energy, and the work fluctuation are respectively derived as
\begin{align}
W_{\mathrm{OIE}}^\mathrm{fin} &= (\omega_{\mathrm{h}} - \omega_{\mathrm{c}})(1 - \xi), \label{wOIEfin}\\
E_{\mathrm{OIE}}^\mathrm{fin} &= \omega_{\mathrm{h}}(1 - \xi),\label{qOIEfin} \\
\delta w_{\mathrm{OIE}}^\mathrm{fin^2} &= (\omega_{\mathrm{h}}^2 + \omega_{\mathrm{c}}^2)(1 - \xi)\xi.\label{dwOIEfin} 
\end{align}
According to Eq.~(\ref{wOIEfin}), because the nonadiabatic transition probability is  bounded by $0 \le \xi < 1/2$, the finite-time OIE can deliver positive  work for  arbitrary driving time $\tau_{\mathrm{dri}}$. This stands in  contrast to the  finite-time OHE, which is restricted to operate  within the regime $\tau_{\mathrm{dri}} > \tau_0$. 

The average measurement times of the CMD1 and CMD2 are $\tau_M^{\mathrm{c}}=\tau p_\mathrm{c}^+/p_\tau(-|+)$ and $\tau_M^{\mathrm{h}}={\tau p_\mathrm{h}^-}/{p_\mathrm{h}^+(1-e^{-\Gamma_\mathrm{h}\tau})}$, respectively (see Appendix \ref{app1}).
Then, the output power  of the finite-time OIE can be determined $P_{\mathrm{OIE}} = {W_{\mathrm{OIE}}^\mathrm{fin}}/{(2\tau_{\mathrm{dri}} + \tau_{\mathrm{h}} + \tau_{\mathrm{c}} + \tau_{\mathrm{M}}^{\mathrm{h}} + \tau_{\mathrm{M}}^{\mathrm{c}})}.$
Because both $P_{\mathrm{OHE}}$ and $P_{\mathrm{OIE}}$ are continuous functions of $\tau_{\mathrm{dri}}$, and given that $P_{\mathrm{OHE}} < 0 < P_{\mathrm{OIE}}$ for $\tau_{\mathrm{dri}} < \tau_0$, there necessarily exists a finite  interval $\epsilon$, allowing the output power of the OIE to exceed that of the OHE in the regime  $\tau_{\mathrm{dri}} < \tau_0 + \epsilon$. 

Furthermore, based on Eq. (\ref{wOIEfin}), when $\tau_\mathrm{dri}\to 0$, where $\xi=0.5$, the minimal work output $(\omega_{\mathrm{h}} - \omega_{\mathrm{c}})/{2}>0$ of the finite-time OIE is reached. In this sudden-quench limit, the power  of the OIE can be expressed as $P_{\mathrm{OIE}}={(\omega_{\mathrm{h}} - \omega_{\mathrm{c}})}/{[2(\tau_{\mathrm{h}} + \tau_{\mathrm{c}} + \tau_{\mathrm{M}}^{\mathrm{h}} + \tau_{\mathrm{M}}^{\mathrm{c}})]}$. Because $I_{\tau}^{\mathrm{c}}, I_{\tau}^{\mathrm{h}} \approx 0$ when $\tau = \tau_{\mathrm{c}} > \tau_{\mathrm{h}}$, we can assume $\tau \le \tau_{\mathrm{c}}$. Consequently, once the energy level gaps $\omega_{\mathrm{c}}$ and $ \omega_{\mathrm{h}}$ and the inverse reservoir temperatures $\beta_{\mathrm{c}} $ and $\beta_{\mathrm{h}}$ are fixed where $N_\mathrm{c,h}$ and $p_{c,h}^{k,j}$ are determined, the output power of the OIE is governed exclusively by  $\gamma_0$.  For example, setting $\tau = \tau_{\mathrm{c}}$ in the  limit $\tau_{\mathrm{dri}} \to 0$, the output power is expressed as
\begin{equation}
P_{\mathrm{OIE}} = \gamma_0 \frac{(\omega_{\mathrm{h}} - \omega_{\mathrm{c}})}{\frac{40}{1 + 2N_{\mathrm{h}}} + \frac{40}{1 + 2N_{\mathrm{c}}} [1 + e^{-\beta_\mathrm{c}\omega_\mathrm{c}}+e^{\beta_\mathrm{h}\omega_\mathrm{h}}]},
\end{equation}
which increases  monotonically with $\gamma_0$. Here, we have assumed relaxation conditions to be $\Gamma_\mathrm{c}\tau_\mathrm{c},\Gamma_\mathrm{h}\tau_\mathrm{h}=20$.

In Eq.~(\ref{dwOIEfin}), the fluctuations in the finite-time OIE originate entirely from nonadiabatic quantum transitions and grow with increasing $\xi$. In the  OHE, however, the work fluctuations stem from both  thermal fluctuations and nonadiabatic quantum transitions. Consequently, the work fluctuations of the OIE are strictly smaller than that of the  OHE.
The analytical proof  as follows. Let  $F(\xi) \equiv \delta w_{\mathrm{OHE}}^\mathrm{fin^2} -\delta w_{\mathrm{OIE}}^\mathrm{fin^2}= F_1 \xi^2 + F_2 \xi + F_3$, where
$F_1 = \omega_{\mathrm{c}}^2 (1 - 4\langle n_{\mathrm{h}} \rangle^2) + \omega_{\mathrm{h}}^2 (1 - 4\langle n_{\mathrm{c}} \rangle^2), 
F_2 = 2 (1 - 2\langle n_{\mathrm{h}} \rangle^2 - 2\langle n_{\mathrm{c}} \rangle^2)\omega_{\mathrm{c}}\omega_{\mathrm{h}} - F_1,$ and
$F_3 = \delta w_\mathrm{OHE}^\mathrm{qs^{2}}.$
Since $\langle n_{\mathrm{h}} \rangle, \langle n_{\mathrm{c}} \rangle \in (-1/2, 0)$,  leading  $F_1 > 0$, and 
$F_2^2 - 4F_1 F_3 = -(1 - 4\langle n_{\mathrm{h}} \rangle^2)(1 - 4\langle n_{\mathrm{c}} \rangle^2)(\omega_{\mathrm{h}}^2 - \omega_{\mathrm{c}}^2)^2 < 0$,  $F(\xi)$ is positive for all $\xi \in [0, 0.5]$,  demonstrating that $w_{\mathrm{OIE}}^\mathrm{fin^2} < w_{\mathrm{OHE}}^\mathrm{fin^2}$. Finally, to achieve zero work fluctuations in the finite-time regime, the OIE can be assisted by counterdiabatic driving (shortcuts to adiabaticity), as elaborated in the Discussion section below.

Combining Eqs.~(\ref{wOIEfin}) and (\ref{qOIEfin}), the finite-time thermodynamic efficiency of the OIE is determined as
$\eta_{\mathrm{OIE}}^\mathrm{fin} = {\eta_\mathrm{OHE}^\mathrm{qs}}/{[1+(I_{\mathrm{c}}+I_{\mathrm{h}})/(\beta_\mathrm{c}E_{\mathrm{OIE}}^\mathrm{fin})]}.$
Because $\eta_{\mathrm{OIE}}^\mathrm{fin}$ is a continuous function of $\xi$, and  the OIE can surpass the Carnot efficiency  in the quasistatic limit ($\xi = 0$), there  exists a finite range of $\xi$ wherein  the relation $\eta_{\mathrm{OIE}}^\mathrm{fin}  > \eta_{\mathrm{C}}$ can be satisfied.

\section{Numerical Analysis}
\begin{figure*}
    \centering
\includegraphics[width=0.8\linewidth]{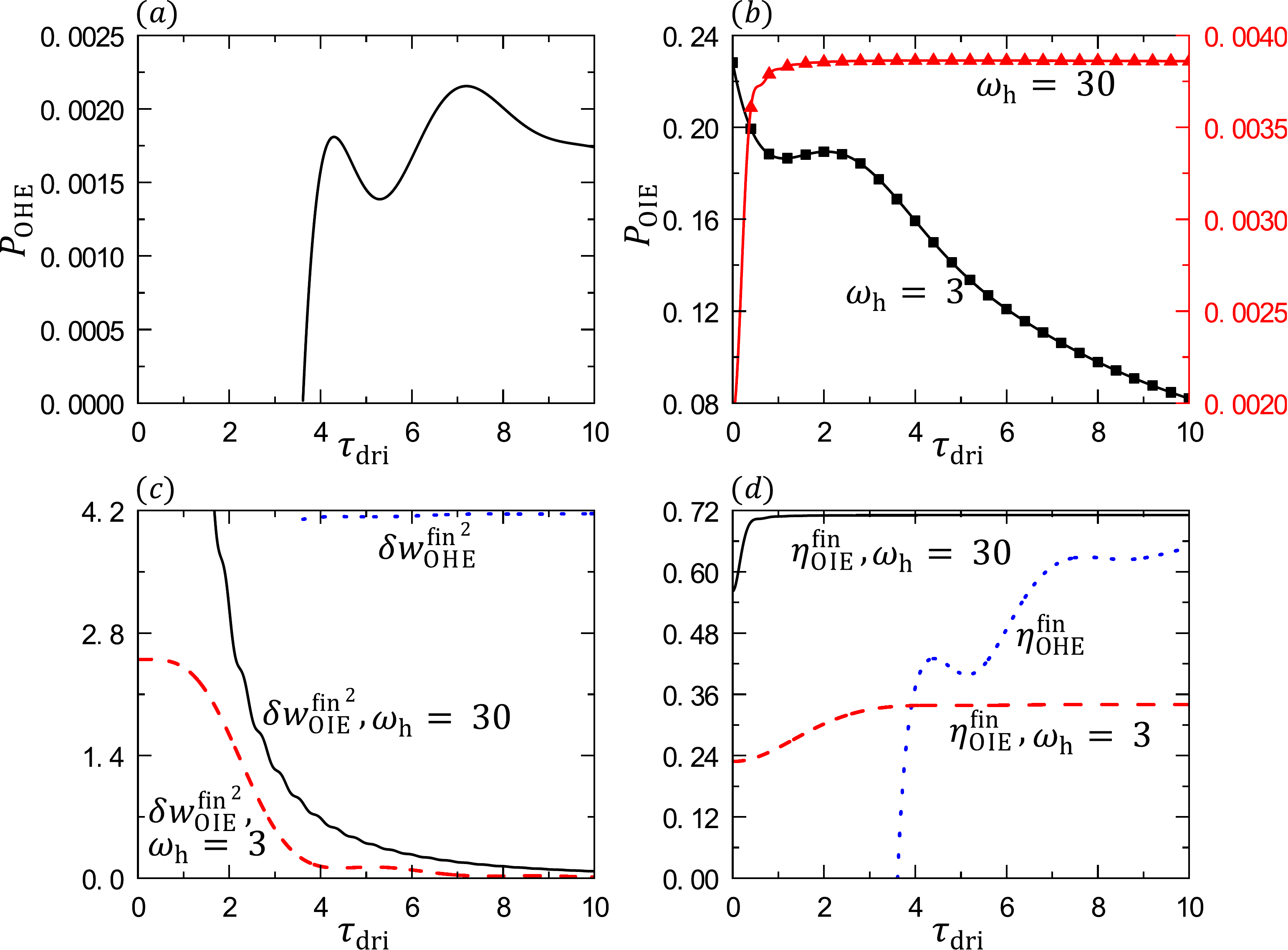}
    \caption{The power (a) and (b), work  fluctuations (c) and the efficiency (d) of the OHE and OIE as functions of the driving time $\tau_\mathrm{dri}$. 
    The energy gaps are $\omega_\mathrm{c}=1$ and $\omega_\mathrm{h}=3$ where $\eta_\mathrm{OHE}^\mathrm{qs}=0.67$. The inverse temperatures are $\beta_\mathrm{c}=1$ and  $\beta_\mathrm{h}=0.3$ where $\eta_\mathrm{C}=0.7$. The  spontaneous emission rate is $\gamma_0=20$. Under these parameters, the relaxation time during the two isochorics stroke and the average measurement time of two CMDs are  $\tau_{\mathrm{h}} + \tau_{\mathrm{c}} = 1.77$ and $\tau_{\mathrm{M}}^{\mathrm{c}} + \tau_{\mathrm{M}}^{\mathrm{h}} = 2.61$. This indicates that the minimum running time of the OIE, $\tau_{\mathrm{h}} + \tau_{\mathrm{c}}+\tau_{\mathrm{M}}^{\mathrm{c}} + \tau_{\mathrm{M}}^{\mathrm{h}}=4.38$, is less than the minimum running time of the OHE, $\tau_{\mathrm{h}} + \tau_{\mathrm{c}}+2\tau_0=8.99$ with $\tau_0=3.61$.}
    \label{pfin}
\end{figure*}

In Figs.~\ref{fig}(a) and (b), we plot the work  of the  OHE and the OIE as functions of the energy level gaps $\omega_{\mathrm{c}}$ and $\omega_{\mathrm{h}}$, respectively. As illustrated in Fig.~\ref{fig}(a), under the given temperatures, the OHE generates non-negligible positive work only within the regime $1 < \omega_{\mathrm{c}} < 8$ and $\omega_{\mathrm{c}} < \omega_{\mathrm{h}} < \beta_{\mathrm{c}}\omega_{\mathrm{c}}/\beta_{\mathrm{h}}$. Furthermore, $W_{\mathrm{OHE}}^{\mathrm{qs}}$ exhibits a nonmonotonic behavior with respect to $\omega_{\mathrm{c}}$ and $\omega_{\mathrm{h}}$, initially increasing and subsequently decreasing, and reaches its  maximum at $\partial W_{\mathrm{OHE}}^{\mathrm{qs}}/\partial \omega_{\mathrm{c}} = 0$ and $\partial W_{\mathrm{OHE}}^{\mathrm{qs}}/\partial \omega_{\mathrm{h}} = 0$. In  contrast, Fig.~\ref{fig}(b) shows that the operational requirement for the OIE reduces to $\omega_{\mathrm{h}} > \omega_{\mathrm{c}}$, and its extracted work possesses no  upper bound. Although Fig.~\ref{fig}(b) displays the parameter window $0 < \omega_{\mathrm{c},\mathrm{h}} < 15$, the physical energy level gaps can extend over $0 < \omega_{\mathrm{c},\mathrm{h}} < \infty$. Crucially, comparing the work scales of Figs. \ref{fig}(a) and (b), the maximum work output of the OIE is approximately $100$ times that of the  OHE  within the  heat-engine regime $\omega_{\mathrm{c}} < \omega_{\mathrm{h}} < \beta_{\mathrm{c}}\omega_{\mathrm{c}}/\beta_{\mathrm{h}}$, and this enhancement reaches $200$-fold when evaluated over the entire parameter space. Consequently, the incorporation of the  CMDs not only  bolsters the engine's parameter robustness, but also produces an order-of-magnitude amplification in work extraction. Moreover, this enhancement ratio increases unboundedly as the parameter range of $\omega_{\mathrm{c},\mathrm{h}}$ expands.

We do not present the work fluctuations here because the OIE  exhibits zero work fluctuations, which represents the maximally stable limit of a thermodynamic machine. Next, in Figs.~\ref{fig}(c) and \ref{fig}(d),   the efficiency of the OIE is  lower than that of the  OHE within the  regime $\omega_{\mathrm{c}} < \omega_{\mathrm{h}} < \beta_{\mathrm{c}}\omega_{\mathrm{c}}/\beta_{\mathrm{h}}$. However, in the extended regime $\omega_{\mathrm{h}} > \beta_{\mathrm{c}}\omega_{\mathrm{c}}/\beta_{\mathrm{h}}$, the efficiency of the OIE not only surpasses that of the  OHE, but also  exceeds the Carnot efficiency limit.

Although the utilization of the  CMDs introduces an additional operational time  $\tau_{\mathrm{M}}^{\mathrm{c}} + \tau_{\mathrm{M}}^{\mathrm{h}}$, the CMDs  provide an order-of-magnitude enhancement in output power under identical  parameters, as illustrated in Figs.~\ref{pfin}(a) and (b). For example, when  $\omega_{\mathrm{h}} = 3$, the maximum output power of the finite-time OIE is approximately $100$ times that of the finite-time OHE. This dramatic enhancement originates from two  mechanisms: first, the  work output of the OIE substantially surpasses the  work of the OHE; second, the OIE can operate under arbitrarily short driving times ($\tau_{\mathrm{dri}} \to 0$), whereas the  OHE is constrained by a non-negligible lower bound $\tau_0$, resulting that the OIE has shorter operation time than the OHE.

Furthermore, in Fig.~\ref{pfin}(c), the OIE exhibits significantly suppressed work fluctuations under identical operating conditions. For example,  when $\tau_{\mathrm{dri}} > 4$,  the work fluctuations in the OIE decay towards zero, however, the OHE still retains a pronounced fluctuations due to  thermal fluctuations. Then, unfortunately,  the order-of-magnitude power enhancement and superior stability of the OIE come at the expense of a  reduction in  efficiency, as shown in Fig.~\ref{pfin}(d). Crucially, this reduction can be circumvented by expanding the driving energy level gap. As shown in Fig.~\ref{pfin}(d), when $\omega_{\mathrm{h}}=30$, the efficiency of the OIE can  surpass that of the OHE,  and even can  exceed the Carnot limit. Although widening the energy level gap induces the reduction of the power and stability, the OIE  preserves a higher output power and superior stability compared to the OHE, as shown by Figs.~\ref{pfin}(b) and \ref{pfin}(c).

\section{Discussion}
In this work, we first  theoretically identify three limitations of the  spin-1/2 Otto heat engine (OHE): (i)  The dependence of trajectory distribution on energy level gap  induced by the Gibbs distribution results in poor parameter robustness, and bounds the work output; (ii) intrinsic thermal fluctuations inevitably give rise to work fluctuations; and (iii) the  efficiency is  constrained by the Carnot bound. Then, we  demonstrate that by introducing  continuous Maxwell's demon (CMD) into the  OHE, called the Otto information engine (OIE), the distribution can be  decoupled from the level gaps, enabling   work output of any size, vanishing work fluctuations, and defeating Carnot efficiency. Furthermore, in the finite-time regime, the OIE simultaneously achieves a substantially enhanced output power, superior  stability, and efficiencies exceeding the Carnot limit.

Crucially, because the CMD  completely eliminates  thermal fluctuations, the finite-time OIE can also achieve  fluctuation-free work extraction by employing counterdiabatic driving to eliminate nonadiabatic transitions \cite{Berry09, OM2019}. 
{Assuming $\varepsilon_j(t)$ and $|j(t)\rangle$  are the eigenvalues and eigenstates of  Hamiltonian $H_{\mathrm{com}}(t)$, then $H_{\mathrm{com}}(t)|j(t)\rangle=\varepsilon_j(t)|j(t)\rangle$.
In the adiabatic approximation, the evolving state of the system can be written as
$|\phi_j(t)\rangle= \exp[-i\int_0^t \mathrm{d}t' \varepsilon_j(t') - \int_0^t \mathrm{d}t' \langle j(t') | \partial_{t'} j(t') \rangle] |j(t)\rangle.$
To ensure that no transitions occur during the first adiabatic stroke, the unitary operator $U(t)$ of this stroke can be chosen as $U(t)=\sum_j |\phi_j(t)\rangle \langle \phi_j(0)|$. Then, the Hamiltonian $H(t)$ corresponding to the unitary operator $U(t)$ can be determined $H(t)=i [\partial_{t}U(t)]U^{\dagger}(t)=H_\mathrm{com}+H_\mathrm{com}^\mathrm{CD}$, where $H_\mathrm{com}^\mathrm{CD}=i\sum_j[| \partial_{t} j(t) \rangle\langle j(t)|-\langle j(t) |\partial_{t} j(t)\rangle | j(t)\rangle \langle j(t)|]=i\sum_{k\neq j}\sum_{j}|k(t)\rangle\langle k(t)|\partial_{t} j(t)\rangle\langle j(t)|$ with $k(t)$ also being the eigenstates of the Hamiltonian $H_{\mathrm{com}}(t)$. 
Because $\langle k(t)|\partial_{t} j(t)\rangle=\langle k(t)|\partial_t H_\mathrm{com}(t) |j(t)\rangle/[\varepsilon_j(t)-\varepsilon_k(t)]$, 
%Writing the compression Hamiltonian in Eq.~(\ref{Hcom}) as $H_{\mathrm{com}}(t) = \vec{b}(t) \cdot \vec{\sigma}$, where $\vec{b}(t) = \{\omega_{\mathrm{ch}}(t)\cos[\theta(t)], \, 0, \, \omega_{\mathrm{ch}}(t)\sin[\theta(t)]\}$ and $\vec{\sigma} = \{\sigma_x, \sigma_y, \sigma_z\}$,
the  counterdiabatic Hamiltonian $H_{\mathrm{com}}^{\mathrm{CD}}(t)$ is  explicitly given by \cite{OM2019}}
\begin{equation}
H_{\mathrm{com}}^{\mathrm{CD}}(t) %=  \frac{\vec{b}(t) \times \dot{\vec{b}}(t)}{2|\vec{b}(t)|^2} \cdot \vec{\sigma} = -\frac{\dot{\theta}(t)}{2}\sigma_y
= -\frac{3\pi t}{2\tau_{\mathrm{dri}}^2}(1 - \frac{t}{\tau_{\mathrm{dri}}})\sigma_y.
\end{equation}
Similarly, for the expansion stroke $H_{\mathrm{exp}}(t) = H_{\mathrm{com}}(\tau_{\mathrm{dri}} - t)$, the auxiliary field takes the form
\begin{equation}
H_{\mathrm{exp}}^{\mathrm{CD}}(t) = \frac{3\pi t}{2\tau_{\mathrm{dri}}^2}(1 - \frac{t}{\tau_{\mathrm{dri}}})\sigma_y.
\end{equation}
Implementing the full control protocols $H_{\mathrm{com}}(t) + H_{\mathrm{com}}^{\mathrm{CD}}(t)$ and $H_{\mathrm{exp}}(t) + H_{\mathrm{exp}}^{\mathrm{CD}}(t)$ along the respective unitary strokes completely eliminate quantum transition ($\xi = 0$), thereby guaranteeing OIE's zero-fluctuation work output  under arbitrarily fast driving.

Consequently, our model provides a versatile paradigm and theoretical blueprint for engineering unbounded work output, zero work fluctuations,  high-power, and high-efficiency quantum  machines.

Finally, in Appendix \ref{sim}, we perform Monte Carlo simulations using the experimental parameters from Ref. \cite{JP29}, which further confirms that the introduction of the CMDs endows the standard Otto heat engine with enhanced output power, zero work fluctuations.

\begin{acknowledgments}
%\acknowledgments{
 Y. Xiao thanks the support in part by the National Natural Science Foundation of China Grant No. NSFC 12234019.  
\end{acknowledgments}
\nocite{*}

\appendix
\section{The calculation of the information obtained by the continuous Maxwell demon}\label{app1}
To determine the explicit expression for $\eta_{\mathrm{OIE}}^\mathrm{qs}$, we must calculated the  information acquired by the CMDs from the system. We first evaluate the information $I_{\mathrm{c}}$ acquired by CMD1.

If CMD1 detects the system in the ground state $|-\rangle$ upon the  first measurement, only a single measurement is performed. The corresponding  information acquired is $-\ln p_\mathrm{c}^-$,  with probability $p_\mathrm{c}^-$. Differently, if the system is detected in the excited state $|+\rangle$ in the initial measurement, the CMD1 then performs stroboscopic measurements at  time intervals $\tau$ until the ground state $|-\rangle$ is first measured. If the CMD1 first measures the system in the ground state at the $m+1$-th ($m \ge 1$) measurement, then the information obtained by the CMD1 is $-\ln p_c^+ \overbrace{-\ln p_\tau(+|+) \dots -\ln p_\tau(+|+)}^{m-1} - \ln p_\tau(-|+) = -\ln p_m$ with the probability $p_m=p_c^+ \overbrace{p_\tau(+|+) \dots p_\tau(+|+)}^{m-1} p_\tau(-|+) = p_c^+ p_\tau(+|+)^{m-1} p_\tau(-|+)$. Here, $p_\tau(+|+)$ and $p_\tau(-|+)$ represent the conditional transition probabilities for the system to remain in the excited state or transition to the ground state after an evolution duration $\tau$ conditioned on the initial excited state, respectively. Furthermore,  $p_\tau(+|+)$ and $p_\tau(-|+)$ satisfy the normalization condition $p_\tau(+|+)+p_\tau(-|+)=1$.
Then the information  acquired by the CMD1 can be  evaluated as
\begin{equation}\label{Ic}
\begin{aligned}
I_\mathrm{c} &= -p_\mathrm{c}^- \ln p_\mathrm{c}^- - \sum_{m=1}^{\infty} p_m \ln p_m \\
&= -p_\mathrm{c}^- \ln p_\mathrm{c}^- - \sum_{m=1}^{\infty} p_\mathrm{c}^+ p_\tau(+|+)^{m-1} p_\tau(-|+)\\
&\times[ \ln p_\mathrm{c}^+ + (m-1)\ln p_\tau(+|+) + \ln p_\tau(-|+) ] \\
&= -p_\mathrm{c}^- \ln p_\mathrm{c}^- - p_\mathrm{c}^+ \ln p_\mathrm{c}^+ \sum_{m=1}^{\infty} [p_\tau(+|+)^{m-1} - p_\tau(+|+)^m] \\
&- p_\mathrm{c}^+ p_\tau(-|+) \ln p_\tau(+|+) \sum_{m=1}^{\infty} (m-1) p_\tau(+|+)^{m-1} \\
& - p_\mathrm{c}^+ \ln p_\tau(-|+) \sum_{m=1}^{\infty} [p_\tau(+|+)^{m-1} - p_\tau(+|+)^m] \\
&= -p_\mathrm{c}^- \ln p_\mathrm{c}^- - p_\mathrm{c}^+ \ln p_\mathrm{c}^+ - p_\mathrm{c}^+ \frac{p_\tau(+|+)}{p_\tau(-|+)} \ln p_\tau(+|+) \\
&- p_\mathrm{c}^+ \ln p_\tau(-|+).
\end{aligned}
\end{equation}

Assuming that $\gamma_0$ represents the spontaneous emission rate and $N_{\mathrm{c}} = (e^{\beta_{\mathrm{c}} \omega_{\mathrm{c}}} - 1)^{-1}$ denotes the mean photon number of the cold reservoir, the rates for the system to absorb and emit a photon during its thermal coupling with the reservoir are given by $\gamma_{\mathrm{c}}^\uparrow = \gamma_0 N_{\mathrm{c}} $ and $ \gamma_{\mathrm{c}}^\downarrow = \gamma_0 (1 + N_{\mathrm{c}})$, respectively. Then the conditional probabilities $p_\tau(+|+) $ and $p_\tau(-|+) $ satisfy the  master equation:
\begin{equation}\label{mas}
\begin{aligned}
\frac{d p_\tau(+|+)}{d\tau} &= p_\tau(-|+)\gamma_\mathrm{c}^\uparrow - p_\tau(+|+)\gamma_\mathrm{c}^\downarrow \\[2ex]
\frac{d p_\tau(-|+)}{d\tau} &= p_\tau(+|+)\gamma_\mathrm{c}^\downarrow - p_\tau(-|+)\gamma_\mathrm{c}^\uparrow.
\end{aligned}
\end{equation}
Subject to the initial condition $p_0(+|+) = 1$ and $p_0(-|+) = 0$, by solving Eq. (\ref{mas}),  $p_\tau(+|+) $ and $p_\tau(-|+) $ can be  obtained:
\begin{equation}\label{sol}
\begin{aligned}
p_\tau(+|+) &= \frac{\gamma_\mathrm{c}^\uparrow}{\gamma_\mathrm{c}^\uparrow + \gamma_\mathrm{c}^\downarrow} + \frac{\gamma_\mathrm{c}^\downarrow}{\gamma_\mathrm{c}^\uparrow + \gamma_\mathrm{c}^\downarrow} e^{-(\gamma_\mathrm{c}^\uparrow + \gamma_\mathrm{c}^\downarrow)\tau} = p_\mathrm{c}^+ + p_\mathrm{c}^- e^{-\Gamma_\mathrm{c} \tau} \\
p_\tau(-|+)& = \frac{\gamma_\mathrm{c}^\downarrow}{\gamma_\mathrm{c}^\uparrow + \gamma_\mathrm{c}^\downarrow} [ 1 - e^{-(\gamma_\mathrm{c}^\uparrow + \gamma_\mathrm{c}^\downarrow)\tau} ] = p_\mathrm{c}^- (1 - e^{-\Gamma_\mathrm{c} \tau}).
\end{aligned}
\end{equation}
Here,
 $\Gamma_\mathrm{c} = \gamma_\mathrm{c}^\uparrow + \gamma_\mathrm{c}^\downarrow$ represents the total effective transition rate associated with the cold reservoir.

 Substituting Eq.~(\ref{sol}) into Eq.~(\ref{Ic}), the acquired information $I_{\mathrm{c}}$ can be determined
\begin{equation}
\begin{aligned}
I_\mathrm{c} &= p_\mathrm{c}^- \ln p_\mathrm{c}^- - p_\mathrm{c}^+ \ln p_\mathrm{c}^+ - p_\mathrm{c}^+ [\frac{p_\mathrm{c}^+}{p_\mathrm{c}^-} + \frac{e^{-\Gamma_\mathrm{c} \tau}}{p_\mathrm{c}^-(1 - e^{-\Gamma_\mathrm{c} \tau})}]\\
&\times[\ln p_\mathrm{c}^+ + \ln(1 + \frac{p_\mathrm{c}^- e^{-\Gamma_\mathrm{c} \tau}}{p_\mathrm{c}^+})]\\ 
&- p_\mathrm{c}^+ [\ln p_\mathrm{c}^- 
 + \ln(1 - e^{-\Gamma_\mathrm{c} \tau})] \\
&= -p_\mathrm{c}^+ \ln p_\mathrm{c}^+ - p_\mathrm{c}^- \ln p_\mathrm{c}^- - p_\mathrm{c}^+ \frac{p_\mathrm{c}^+}{p_\mathrm{c}^-} \ln p_\mathrm{c}^+ \\
&- p_\mathrm{c}^+ \frac{e^{-\Gamma_\mathrm{c} \tau}}{p_\mathrm{c}^-(1 - e^{-\Gamma_\mathrm{c} \tau})} \ln p_\mathrm{c}^+ \\
&- p_\mathrm{c}^+ \frac{p_\mathrm{c}^+ + p_\mathrm{c}^- e^{-\Gamma_\mathrm{c} \tau}}{p_\mathrm{c}^-(1 - e^{-\Gamma_\mathrm{c} \tau})} \ln(1 + \frac{p_\mathrm{c}^- e^{-\Gamma_\mathrm{c} \tau}}{p_\mathrm{c}^+})\\ 
&-  p_\mathrm{c}^+ \ln p_\mathrm{c}^- - p_\mathrm{c}^+ \ln(1 - e^{-\Gamma_\mathrm{c} \tau}) 
= I_\mathrm{min}^\mathrm{c} + I_\tau^\mathrm{c}, 
\end{aligned}
\end{equation}
where
 $I_\mathrm{min}^\mathrm{c} = -p_\mathrm{c}^+ \ln p_\mathrm{c}^+ - p_\mathrm{c}^- \ln p_\mathrm{c}^- - p_\mathrm{c}^+ {p_\mathrm{c}^+}/{p_c^-} \ln p_c^+ - p_\mathrm{c}^+ \ln p_\mathrm{c}^- = -\ln p_\mathrm{c}^- - {p_\mathrm{c}^+}/{p_\mathrm{c}^-}\ln p_\mathrm{c}^+$ and $I_\tau^\mathrm{c} = -p_\mathrm{c}^+ {e^{-\Gamma_\mathrm{c} \tau}}/[p_\mathrm{c}^-(1-e^{-\Gamma_\mathrm{c} \tau})] \ln p_\mathrm{c}^+ - p_\mathrm{c}^+ {(p_\mathrm{c}^+ + p_\mathrm{c}^- e^{-\Gamma_\mathrm{c} \tau})}\ln ( 1 + {p_\mathrm{c}^- e^{-\Gamma_\mathrm{c} \tau}}/{p_\mathrm{c}^+} )/[p_\mathrm{c}^-(1-e^{-\Gamma_\mathrm{c} \tau})]  - p_\mathrm{c}^+ \ln(1 - e^{-\Gamma_\mathrm{c} \tau})$.

Following the same procedure as the CMD1, the average information $I_{\mathrm{h}}$ acquired by the CMD2 can be  established
\begin{equation}
I_{\mathrm{h}} =I_\mathrm{min}^\mathrm{h} + I_\tau^\mathrm{h},
\end{equation}
where 
$I_\mathrm{min}^\mathrm{h}=-\ln p_\mathrm{h}^+-p_\mathrm{h}^-/p_\mathrm{h}^+\ln p_\mathrm{h}^-$
represents the minimal information acquired by the CMD2, and $I_\tau^\mathrm{h}= -p_\mathrm{h}^- {e^{-\Gamma_\mathrm{h} \tau}}/{[p_\mathrm{h}^+(1-e^{-\Gamma_\mathrm{h} \tau})]} \ln p_\mathrm{h}^- - p_\mathrm{h}^- {[p_\mathrm{h}^- + p_\mathrm{h}^+ e^{-\Gamma_\mathrm{h} \tau}]}\ln (1 + {p_\mathrm{h}^+ e^{-\Gamma_\mathrm{h} \tau}}/{p_\mathrm{h}^-} )/{[p_\mathrm{h}^+(1-e^{-\Gamma_\mathrm{h} \tau})]}  - p_\mathrm{h}^- \ln(1 - e^{-\Gamma_\mathrm{h} \tau})$ with $\Gamma_{\mathrm{h}} \equiv \gamma_0 (2N_{\mathrm{h}} + 1)$ being the total transition rate, where $N_{\mathrm{h}} = (e^{\beta_{\mathrm{h}} \omega_{\mathrm{h}}} - 1)^{-1}$ is the mean photon number of the hot reservoir and we have assumed identical spontaneous emission rates $\gamma_0$ for both reservoirs.

To determine the output power of the OIE, the measurement time consumed by both CMD1 and CMD2  must be quantified. Here, we first evaluate the average measurement time required by the CMD1. If the system is detected in the ground state upon the first measurement, no additional waiting time is incurred. If the ground state is first registered at the $m+1$-th measurement, the elapsed duration is $m\tau$. Then, the   average measurement time $\tau_M^{\mathrm{c}}= \tau\sum_{m=1}^{\infty} m p_m$  can be obtained 
\begin{equation}
\begin{aligned}
\tau_M^{\mathrm{c}}=\frac{p_\mathrm{c}^+p_\tau(-|+)}{p_\tau(+|+)}\sum_m m p_\tau(+|+)^m=
\frac{\tau p_\mathrm{c}^+}{p_\tau(-|+)}.
\end{aligned}
\end{equation}
Similarly, the average measurement time of the CMD2 reads
\begin{equation}\label{th}
\tau_M^{\mathrm{h}}=\frac{\tau p_\mathrm{h}^-}{p_\mathrm{h}^+(1-e^{-\Gamma_\mathrm{h}\tau})},
\end{equation}

\section{The simulations of the OIE with the experimental parameters}\label{sim}

\begin{figure}
    \centering
\includegraphics[width=1\linewidth]{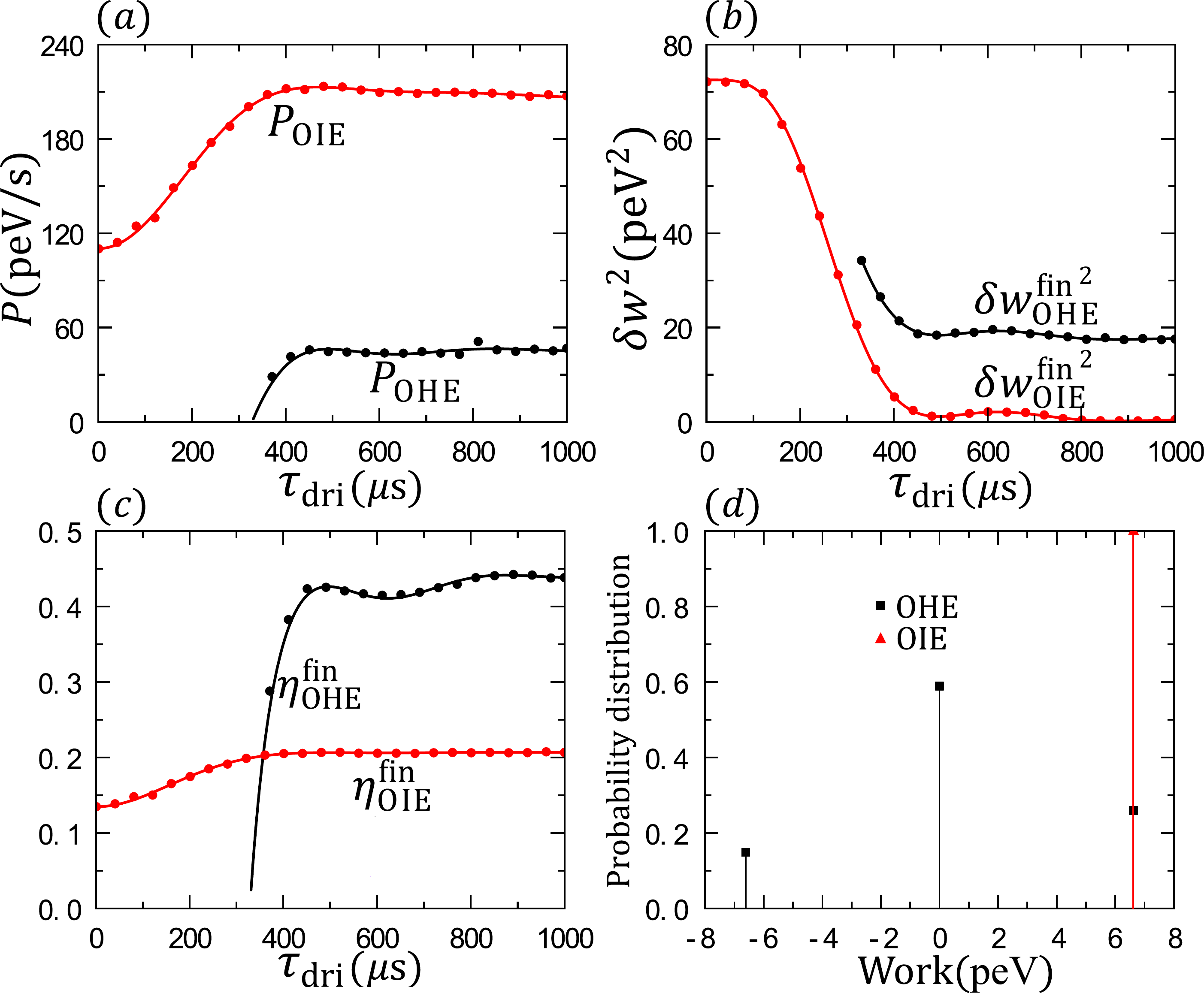}
    \caption{The power (a), work  fluctuations (b) and the efficiency (c) of the OHE and OIE as functions of the driving time $\tau_\mathrm{dri}$. (d) The work distributions of the OHE and OIE with  $\tau_\mathrm{dri}=1\mu 
    \mathrm{s}$, under the counterdiabatic driving regime.
    The energy gap are $h \omega_\mathrm{c}\approx8.27 \mathrm{peV}$ and $h \omega_\mathrm{h}\approx14.89 \mathrm{peV}$. The temperatures of the cold and hot reservoirs are $k_B T_\mathrm{c}= 6.6 \mathrm{peV}$ and $k_B T_\mathrm{h}= 21.5 \mathrm{peV}$. In (a)-(c), solid circles and solid lines represent simulated  and theoretical values, respectively.}
    \label{sum}
\end{figure}
To test our theoretical framework under realistic physical conditions, we perform Monte Carlo trajectory simulations using the realistic physical parameters of the liquid-state nuclear magnetic resonance (NMR) experiment reported in Ref. \cite{JP29}. The working medium is the $^{13}\text{C}$ nuclear spin driven by resonant radio-frequency fields, and  the frequency fields are  $\omega_\mathrm{c}= 2.0\text{ kHz}$ and $\omega_\mathrm{h} = 3.6\text{kHz}$. The effective spin temperatures of the cold and hot reservoirs are chosen as $k_B T_1 = 6.6\text{ peV}$ and $k_B T_2 = 21.5\text{ peV}$, respectively. The thermalization duration is governed by the scalar coupling with the $^1\text{H}$ nucleus on the timescale of $\sim 7\text{ ms}$. 

For each  cycle, the system state transitions are evaluated stochastically along the quantum trajectories. The  nonadiabatic transition probability $\xi$ is determined by Eq. (\ref{Hcom}). In the standard OHE, the initial state of the system during the adiabatic strokes are sampled from the Gibbs distributions. In contrast, for the OIE, the system is stroboscopically monitored by CMD1 and CMD2 with time intervals $\tau_{\text{meas}}=7\text{ms}$ until the ground state $\vert{}-\rangle$ and excited state $\vert{}+\rangle$ are respectively detected. The  stochastic information acquired $I_\mathrm{c,h}$  and the measurement time $\tau_M^\mathrm{c,h}$ are accumulated along each trajectory following  Eqs. (\ref{Ic})–(\ref{th}).

The simulated power, work fluctuations and efficiency are shown in Fig. \ref{sum}, and are in  agreement with our theoretical predictions. As shown  by Fig. \ref{sum}, compared with the finite-time OHE, the finite-time OIE can achieve a multiplication in both output power and stability, and can even attain zero work fluctuations, with the sacrifice of the efficiency, as predicted by our results. However, the OHE's efficiency can be enhanced by operating  beyond the Carnot boundary, as determined in Eq. (\ref{etaOIE}).

\bibliography{ref}

\end{document}